\documentclass[preprint]{aastex}
\newcommand{\pmn}{PMN~J1838--3427}

\slugcomment{}

\begin{document}

\title{
\pmn: A new gravitationally lensed quasar\footnote{
Based on observations using the Very Large Array (VLA)
and Very Long Baseline Array (VLBA) of the National Radio
Astronomy Observatory (NRAO), the NASA/ESA Hubble Space
Telescope (HST), the 3.6m telescope of the European Southern
Observatory (ESO) at La Silla, the du Pont telescope
at Las Campanas Observatory (LCO),
and the Australia Telescope Compact Array (ATCA).
The NRAO is a facility of the National Science
Foundation (NSF) operated under cooperative agreement by
Associated Universities, Inc.\ The HST data were obtained
from the Space Telescope Science Institute, which is operated
by AURA, Inc., under NASA contract NAS~5-26555.
ATCA is part of the Australia Telescope which is funded by the
Commonwealth of Australia for operation as a National Facility
managed by CSIRO.
}
}

\author{Joshua N.\ Winn\altaffilmark{2},
Jacqueline N.\ Hewitt\altaffilmark{2},
Paul L.\ Schechter\altaffilmark{2},
Alan Dressler\altaffilmark{3},
E.E.\ Falco\altaffilmark{4},
C.D.\ Impey\altaffilmark{5},
C.S.\ Kochanek\altaffilmark{4},
J.\ Leh\'{a}r\altaffilmark{4},
J.E.J.\ Lovell\altaffilmark{6},
B.A.\ McLeod\altaffilmark{4},
Nicholas D.\ Morgan\altaffilmark{2},
J.A.\ Mu\~{n}oz\altaffilmark{7},
H.-W.\ Rix\altaffilmark{8},
Maria Teresa Ruiz\altaffilmark{9}
}

\altaffiltext{2}{Department of Physics, Massachusetts Institute of Technology,
    Cambridge, MA 02139}
\altaffiltext{3}{The Observatories of the Carnegie Institution of Washington,
    813 Santa Barbara St., Pasadena, CA 91101}
\altaffiltext{4}{Harvard-Smithsonian Center for Astrophysics, 60 Garden St.,
    Cambridge, MA 02138}
\altaffiltext{5}{Steward Observatory, University of Arizona, Tucson, AZ 85721}
\altaffiltext{6}{Australia Telescope National Facility, CSIRO, PO Box 76,
    Epping, NSW 1710, Australia}
\altaffiltext{7}{Instituto de Astrof\'{\i}sica de Canarias,
    E-38200 La Laguna, Tenerife, Spain}
\altaffiltext{8}{Max-Planck-Institut f\"{u}r Astronomie, Koenigstuhl 17, Heidelberg,
    D-69117, Germany}
\altaffiltext{9}{Departamento de Astronom\'{\i}a, Universidad de Chile, Casilla 36-D,
    Santiago, Chile}

\begin{abstract}
We report the discovery of a new double-image quasar that
was found during a search for gravitational lenses in the southern sky.
Radio source \pmn~is composed of two flat-spectrum components
with separation $1\farcs00$, flux density ratio 14:1 and
matching spectral indices, in VLA and VLBA images.
Ground-based $BRI$ images show the optical counterpart
(total $I=18.6$) is also double with the same separation
and position angle as the radio components. An HST/WFPC2 image
reveals the lens galaxy. The optical flux ratio (27:1) is higher
than the radio value probably due to differential extinction
of the components by the lens galaxy. An optical spectrum of the bright
component contains quasar emission lines at $z=2.78$ and several
absorption features, including prominent Ly-$\alpha$ absorption.
The lens galaxy redshift could not be measured but is
estimated to be $z=0.36 \pm 0.08$. The image configuration
is consistent with the simplest plausible models for the lens potential.
The flat radio spectrum and observed variability of \pmn~suggest
the time delay between flux variations of the components is
measurable, and could thus provide an independent
measurement of $H_{0}$.
\end{abstract}


\section{Introduction}
\label{sec:intro}

The new double-image quasar presented in this paper is the
first to result from a program that three of us
(J.N.W., J.N.H., and P.L.S.)
began to find new radio-loud gravitational lenses in
the southern sky. Our main goal is to find suitable lenses
for time-delay measurements.
The time delays between the flux variations of the multiple images
of a gravitationally lensed quasar can be used as a ``one-step'' measurement
of the Hubble constant $H_{0}$ \citep{refsdal} or, more precisely,
a combination of angular-diameter distances between the Earth,
lens galaxy, and quasar
(for recent reviews see \citet{myers99b,schech97}).
In addition, the measured lensing rate in a well-defined
sample of extragalactic sources places interesting limits
on the cosmological constant (\cite{lambda1,lambda2,lambda3}).

Our search methodology
will be described in a future paper;
we confine ourselves here to a summary.
We selected southern sources
because the southern hemisphere is relatively unexplored
for lenses and therefore more likely
to contain bright and useful specimens.
The southern limit is $\delta = -40\arcdeg$ to permit the use of
the NRAO Very Large Array (VLA) and Very Long Baseline Array (VLBA),
instruments that facilitate the search.

We selected only sources with flat spectral indices
($\alpha \geq -0.5$, where $S_{\nu} \propto \nu^{\alpha}$)
as measured between the
4.85~GHz Parkes-MIT-NRAO catalog \citep{pmn}
and the 1.4~GHz NRAO VLA Sky Survey \citep[NVSS]{nvss}.
Flat-spectrum sources tend to be core-dominated,
and therefore variable (a prerequisite for measuring time delays),
easily recognized when lensed, and easily mapped by an automatic procedure.
In this respect our program is similar to the
Cosmic Lens All-Sky Survey (CLASS; \citet{class}), a program that
(after including the sample
of the Jodrell-VLA Astrometric Survey, JVAS)
has identified at least 15 new lenses
among 15000 northern-hemisphere radio sources.

We observed each object for 30 seconds
at 8.46~GHz with the VLA in its A configuration.
Objects exhibiting multiple compact components
(about 5\% of the sample) were selected as
lens candidates and scheduled for appropriate
follow-up observations, including
multifrequency VLA imaging, VLBA imaging, and optical imaging.
The goal of the follow-up observations
is to determine whether the components have
similar spectral properties and surface
brightnesses (as lensed images should) and to search
for a lens galaxy.

For the particular case of \pmn, the chronology of
observations was as follows. The initial VLA image from
1998 May 19 contained two compact
components separated by $1\farcs00$.
The most likely explanation for this morphology
was a core-jet or core-hotspot structure.
However, in ground-based images obtained during
1999 April 10-15, the optical counterpart was
revealed to be a double with the same separation
and position angle as the radio double, thereby
suggesting \pmn~was either a
binary quasar or a gravitational lens.

On 1999 July 19, multifrequency
observations with the VLA revealed
that the spectral indices
of the components were the same.
This evidence favored the gravitational lens hypothesis,
since lensing is achromatic.
We used the VLBA on 1999 October 11 to search for matching
milliarcsecond substructure within the radio components,
which is characteristic (although not required) of lensed images.
Both components were detected, but
no matching substructure was seen.

On the strength of the evidence thus far, \pmn~was
included in the CfA-Arizona Space Telescope Lens Survey (CASTLES),
an effort by seven of us
(E.E.F., C.D.I., C.S.K., J.L., J.A.M., B.A.M., and H.-W.R.)
to use the {\em Hubble Space Telescope} (HST)
to observe all the known galaxy-scale gravitational lenses.
The HST/WFPC2 image, obtained on 2000 March 1, revealed
a diffuse light source between the quasar components
which is naturally interpreted as a lens galaxy.
This left no doubt that \pmn~is
a gravitationally lensed quasar.

We obtained optical spectra of the quasar and lens galaxy
with the ESO 3.6m telescope
at La Silla on 2000 March 4.
This allowed us to measure the source redshift $z=2.78$
but our lens galaxy spectrum was inconclusive.
Finally, in order to assess the radio variability of \pmn~we
measured its total flux at two frequencies
with the Australia Telescope Compact Array (ATCA)
on 2000 April 5.

Subsequent sections of this paper present these
observations in logical rather than chronological order.
Sections~\ref{sec:vla} through~\ref{sec:atca} present
the radio properties of the system, as revealed
by measurements with the VLA, VLBA, and ATCA.
Sections~\ref{sec:hst} and~\ref{sec:lco}
present space-\ and ground-based
optical images, and \S~\ref{sec:spectra}
presents optical spectra of
the quasar and lens galaxy.
We consider simple models of the lens potential
in \S~\ref{sec:model}.
Finally, in the last section we review the evidence
that \pmn~is a gravitationally lensed quasar
and discuss the prospects for measuring
and interpreting the time delay
between its lensed images.

\section{VLA images}
\label{sec:vla}

\begin{figure}
\figurenum{1}
\plotone{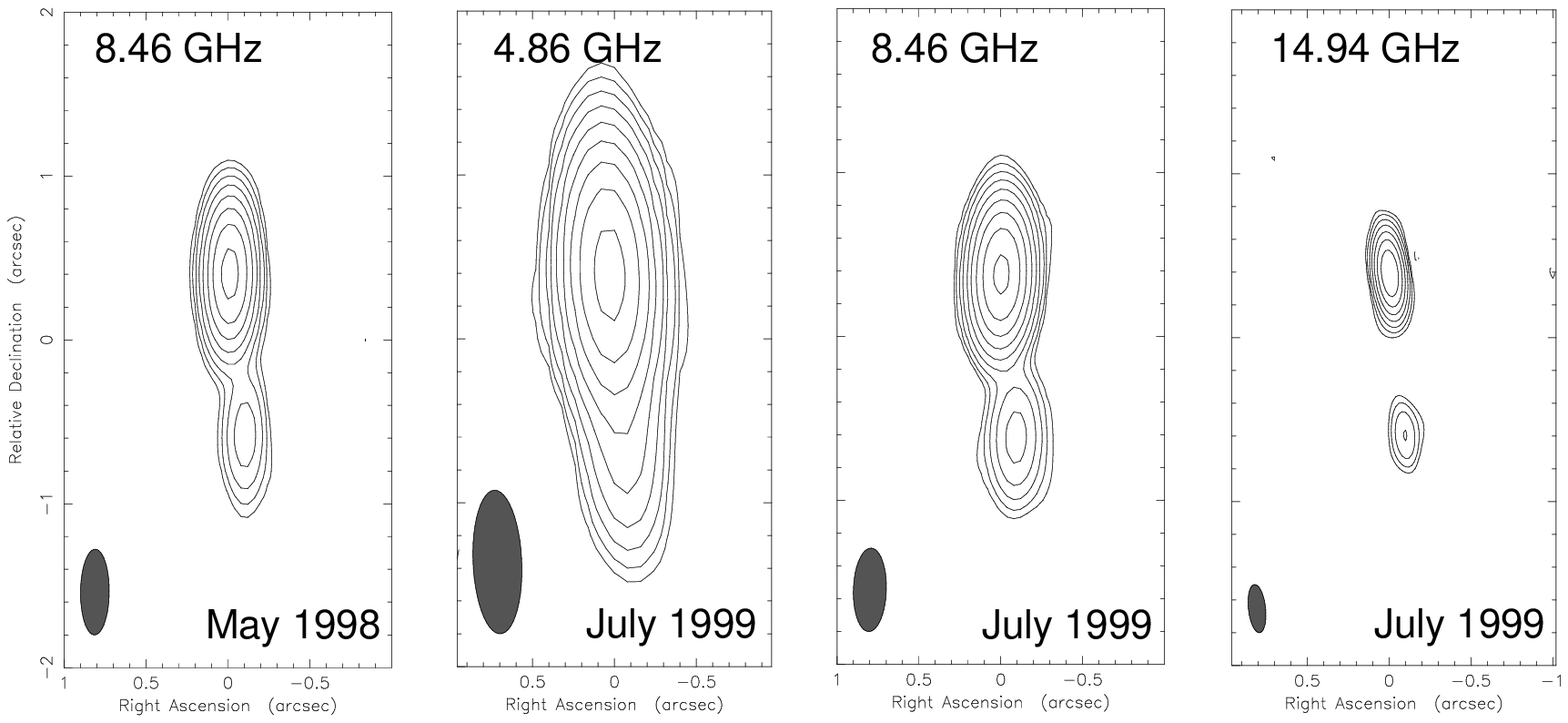}
\caption{
VLA images (2\arcsec $\times$ 4\arcsec) of \pmn.
Contours begin
at $3\sigma$ and increase by powers of 2,
where $\sigma$ is the RMS level in the residual map.
From left to right
$\sigma = 0.35$, 0.20, 0.24, and 0.49 mJy/beam.
The restoring beams (inset in the lower left of each map)
are elongated N/S due to low-elevation observing.
Beam diameters are listed in column 3 of Table~\ref{tbl:vla}.
}
\label{fig:vla}
\end{figure}

The date, frequency, resolution, and sensitivity of
the VLA observations are listed in the first
four rows of Table~\ref{tbl:vla}.
In all cases the VLA was in its A configuration and
the total observing bandwidth was 100 MHz.
Radio source 3C286 was used to set
the absolute flux scale, following the
procedures suggested in the VLA Calibrator
Manual and adopting flux densities 7.49, 5.52, and 3.42 Jy
at 4.86, 8.46, and 14.94~GHz respectively.
Calibration was performed with standard
procedures in AIPS. Deconvolution and imaging
were performed with Difmap \citep{difmap}. This included
phase-only self-calibration with a 30-second
solution interval. The final images,
created with uniform weighting and
an elliptical Gaussian restoring beam,
are shown in Figure~\ref{fig:vla}.

In each image there are two components.
Throughout this paper we refer to the northern
component of \pmn~as A, and the southern
component as B.
To measure the separation and
flux density ratio of A and B,
we fit a two-component model
to the self-calibrated
visibilities using the ``modelfit'' utility of Difmap.
After subtracting the best-fit
model from the data, the residual images were all
consistent with thermal noise.

The best-fit parameters are listed in Table~\ref{tbl:vla}.
The flux density ratios are in rough agreement but are not
equal within the quoted uncertainties. For a gravitational
lens the ratios are expected to be approximately equal,
but one reason to expect small discrepancies
is variability of the background object
(see \S~\ref{sec:summary} for further discussion).
Two-point spectral indices were computed for each component
based on the 1999 July 19 observations. The results are
listed in Table~\ref{tbl:indices}.
Both components are verified
to have flat radio spectra.

The coordinates of component A, based on the
1999~July~19 images, are
R.A.~(J2000)$~=
18^{{\mathrm h}}38^{{\mathrm m}}28^{{\mathrm s}}.50$,
Dec.~(J2000)~$ =
-34\arcdeg 27\arcmin 41\farcs6$,
within $0\farcs2$.

\section{VLBA image}
\label{sec:vlba}

We observed \pmn~for 4 hours with the VLBA
on 1999 October 11.
The St.\ Croix and Brewster antennas were unavailable.
The observing bandwidth was 48 MHz, divided into
8~IFs, with an average
frequency of 4.975~GHz.
The visibility averaging time was 1 second.
This level of spectral and time sampling
was sufficient to prevent significant bandwidth-
and time-average smearing over
the required field of view.

Calibration was performed with standard procedures in AIPS.
Fringe fitting for sources with widely spaced components
is problematic without a prior model of
the source, because the source structure causes rapid
time variation of the solutions.
For this reason we used a two-step procedure.
First, the data were fringe-fitted using a point source
model and a 30-second solution interval.
This allowed a preliminary image to be made,
in which components A and B were both detected.
Then, the calibration was redone by fringe-fitting with
reference to a two-component model based on the
preliminary image and a 3-minute solution
interval. The final images were created with Difmap,
after repeated iterations of model fitting
and phase-only self-calibration.

\begin{figure}
\figurenum{2}
\plotone{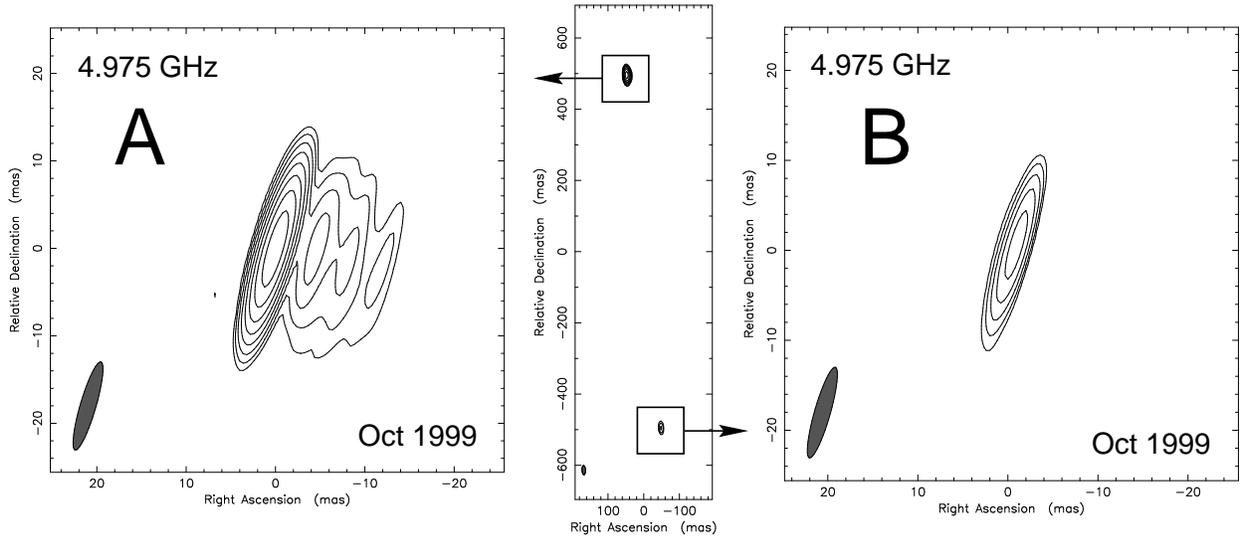}
\caption{
VLBA image of \pmn.
A wide-field image, displaying
both northern and southern components,
is displayed in the center panel.
Close-ups of the northern and southern
components are shown in the left and right
panels, respectively.
Contour levels begin
at $3\sigma$ and increase by powers of 2, where $\sigma$
is the RMS level in the residual map, which
is 0.20 mJy/beam near A and
0.18 mJy/beam near B.
The synthesized beam, depicted in the lower
left of each map, has FWHM diameters
10.5 $\times$ 1.8 mas.
}
\label{fig:vlba}
\end{figure}

The final uniformly-weighted
images are shown in Figure~\ref{fig:vlba}.
Both a wide-field
map and close-ups of each component are displayed.
Component B is unresolved.
Component A is partially resolved; in addition to an unresolved
component there is extended emission to the west,
comprising about 10\% of the total flux density.

A model for the surface brightness distribution
was constructed with Difmap,
by fitting a circular Gaussian to
each compact component (which we continue to
identify as A and B), and then using the Clean
algorithm to account for the diffuse emission
west of A with a large number of point sources.
During the subsequent iterations of model-fitting and
self-calibration, the Clean components
were kept fixed and the circular Gaussian parameters
were varied. The best-fit positions and
flux densities of A and B are reported in the
last entry of Table~\ref{tbl:vla}.
The separation between the Gaussian components
is 996.1 mas and the position angle is $5\fdg59$
east of north. The total flux density of the
diffuse emission is 16.7 mJy.

The flux density of B is
the same in the VLBA and VLA images, but the flux density of A
is 27\% smaller in the VLBA image.
As a result the VLBA flux density ratio (10.6)
is smaller than the VLA ratios (which average 14.6).
Even when the total flux
density of the diffuse emission west of A
is added to that of A, the VLBA ratio is only 11.8.
One possible explanation is that the VLA
image of A includes relatively diffuse flux
that is missing from the VLBA image.
The shortest VLBA baseline (236 km, or 4 M$\lambda$) is longer
than the longest VLA baseline (35 km, or 0.6~M$\lambda$), so
there is a range of spatial frequencies that
are unresolved by the VLA and invisible to the VLBA.
Another possibility to explain part or all of
the discrepancy is that the source varied between observing epochs.
Due to the time delay between components, source
variability would cause fluctuations in the
instantaneous flux density ratio.

Since \pmn~is a gravitational lens, there should
be radio emission east of B corresponding to the
emission west of A. However, given the peak brightness
of the emission west of A (2.5 mJy/beam) and the
magnification ratio described above, one would expect the
brightness of such emission to be at most 0.24 mJy/beam,
which is not much higher than the RMS level of 0.18 mJy/beam.
Confirmation of matching milliarcsecond substructure will
require deeper imaging.

\section{ATCA measurements}
\label{sec:atca}

To test for variability we examined \pmn~with the
Australia Compact Telescope Array (ATCA)
on 2000 April 5, observing simultaneously at 4.80 and 8.64~GHz
while the array was in the 6D configuration.
The observation was divided into three scans of one minute
each, with 2 hours between scans to improve
the $uv$-coverage. Calibration was performed with
the software package MIRIAD, and imaging with Difmap.
The absolute flux density scale was set by observations
of PKS~B1934--638, which is believed to match the 3C286-based
flux scale within~3\%.

The antenna spacings were not large enough to
resolve the components, so we used the VLA models
to fix the separation of the two components
and varied their flux densities to achieve the best fit
to the ATCA data.
Since the individual flux densities
are covariant we report only their sum
in Table~\ref{tbl:fluxes}. The uncertainty
was estimated as the quadrature sum of the range
in total flux obtained by analyzing each
one-minute scan separately,
and a 3\% uncertainty due to absolute
flux calibration.
Table~\ref{tbl:fluxes} also
lists other flux density measurements reported in this
paper and in various published radio catalogs.

The total ATCA flux density at 4.80~GHz was the same
as the most recent 4.86~GHz VLA measurement.
However, the total flux density at 8.64~GHz
was 44\% higher than the most recent 8.46~GHz VLA measurement,
indicating strong variability. (The expected difference
due to spectral index alone is 0.6\%, and the
combined measurement uncertainty is 5\%.)
Variability at 8.46~GHz is corroborated by
the 12\% variation in total flux density observed
between the two VLA measurements.

\section{HST image}
\label{sec:hst}

On 2000 March 1, three dithered
exposures (700 sec, 700 sec, 600 sec) of \pmn~were
acquired during one HST orbit,
using the Wide Field and Planetary Camera 2 (WFPC2)
and the F814W filter.
The data were reduced with the standard CASTLES pipeline:
the exposures were registered, weighted by exposure time,
and combined using a 3-sigma ``ccdclip'' rejection
algorithm. (See \citet{lehar} for other examples of two-image
lenses observed in the CASTLES program.)

\begin{figure}
\figurenum{3}
\plotone{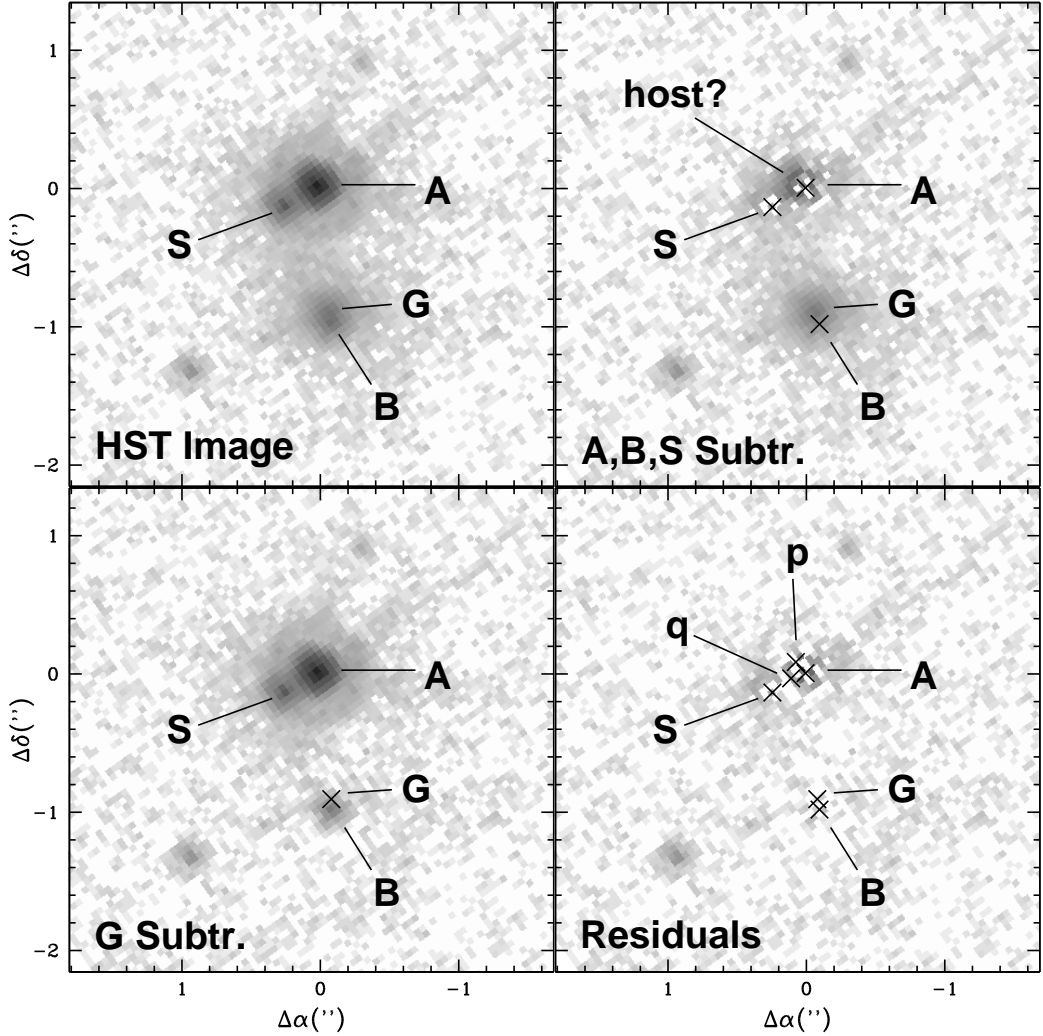}
\caption{
{\em Upper left.}
HST/WFPC2 image of \pmn~through filter F814W
($\approx I$).
Here and in all other panels, the
gray levels are logarithmically scaled.
{\em Upper right.}
Same, but with components A, B and S
subtracted in order to highlight
the lens galaxy G. The residuals
near A are tentatively identified as
due (at least in part)
to the quasar host galaxy.
{\em Lower left.}
Only component G has been
subtracted, in order to highlight
the dim quasar component B.
{\em Lower right.}
All model components
have been subtracted.
There are still residuals near A possibly
due to the host galaxy.
}
\label{fig:hst}
\end{figure}

The upper left panel of figure~\ref{fig:hst} is
a 3\arcsec $\times$ 3\arcsec~subraster of
the final image.
Components A and B are present with the same
separation and position angle as the corresponding
radio components.
Crucially, the southern component is not compact.
There is diffuse emission extending northward from B
towards A, which is labeled G.
The natural interpretation
is that G is a lens galaxy:
it lies along the line joining A and B,
and its position near B is consistent with the large
flux density ratio between A and B
(see \S~\ref{sec:model} for simple models).

In addition, there is a third compact object
$0\farcs28$ southeast of A, labeled S.
Object S is almost certainly not a third
image of the background quasar, because
it is silent in all our radio images.
In particular, the 4.86~GHz image
of 1999 July 19 requires S to be
at least 200 times dimmer than A, whereas in the
optical image it is about 8 times dimmer.
Missing optical images can be plausibly
explained by dust extinction
(e.g.\ B1152+199; \citet{myers99a})
but missing radio images
defy simple explanation.

The low galactic latitude ($b=-12\arcdeg$) of the
field suggests that S is a foreground star.
To estimate the {\em a posteriori} probability of a
star intruding on the lens, we estimated the local
mean density of stars by counting all
the stars in the PC field (32\arcsec $\times$ 32\arcsec)
that are at least as bright as S.
Based on this mean density,
the probability that at least one star
would appear within $0\farcs5$ of
component A, component B, or the line joining them,
is 5\%. This is somewhat low but within reason.

To measure the relative positions and magnitudes of
A, B, G, and S, we created parameterized
models of the image using a twice-oversampled
PSF created by TinyTim v4.4 \citep{tinytim}.
The models were fit to the data using the
procedures of \citet{lehar}.
The first model contained three point sources (representing
A, B, and S) and a circular de Vaucouleurs profile (G).
The residual image contained a diffuse pattern of
residuals centered on component A with an
integrated flux equal to 12\% of the combined
flux of A and S, suggesting
that A is not adequately
modeled by a point source.
Possibly, the host galaxy of the background quasar
and/or additional foreground objects
are contributing light.

We tried accounting for this
extra light with additional
model components,
such as a circular de Vaucouleurs profile,
an elliptical de Vaucouleurs profile, 
and extra point sources.
There is no compelling reason to recommend
one of these models over the others,
but we judged that the residuals appeared
most random for a model with
two extra point sources.
We refer to these two extra point sources
as p and q.
The best-fit parameters
of the model are listed in Table~\ref{tbl:hst}.
Parameter uncertainties were estimated
in the same manner as \citet{lehar}: three
terms were added in quadrature, representing
the statistical error in the fit, the range
in parameters obtained with different choices of
the model PSF, and the range obtained
with two different modeling programs.

The upper right panel of Figure~\ref{fig:hst}
shows the image after components A, B and S
of the best-fit model have been subtracted.
This allows the lens galaxy G and the excess
residuals near A to be seen clearly.
In the lower left panel, only G has been
subtracted, highlighting component B.
It is worth noting that the A/B flux ratio in
this image (27:1) is
much higher than any of the radio flux density ratios.
This is probably due to the proximity of B
and G. Optical extinction due to the
lens galaxy should be greater for component B than for A.

In the lower right panel the entire model
(including p and q) has been subtracted.
There are still residuals near A, extending
in both directions nearly perpendicular to
the A/B separation.
The elongation of the residuals is suggestive
of tangentially-stretched emission from
a host galaxy.
Deeper HST or adaptive-optics
imaging in the infrared will be useful
to clarify the nature of the putative host galaxy.

To connect the WFPC2 photometry
to the Johnson-Kron-Cousins system 
we approximated F814W $\approx I$
and adopted a zero-point magnitude
of 21.69, as did \citet{lehar}. This
zero-point is based on the calibration
of \citet{holtzmann},
but with a gain of 7 and a correction
of 0.1 mag for finite aperture.
The resulting magnitude of
component A is $I = 19.21$,
and the total magnitude of all
components is $I = 18.79$,
with a scatter of 0.05 between different
models and an additional uncertainty
of at least 0.05 due to the choice of zero-point
magnitude.

\section{Ground based optical images}
\label{sec:lco}

\begin{figure}
\figurenum{4}
\plotone{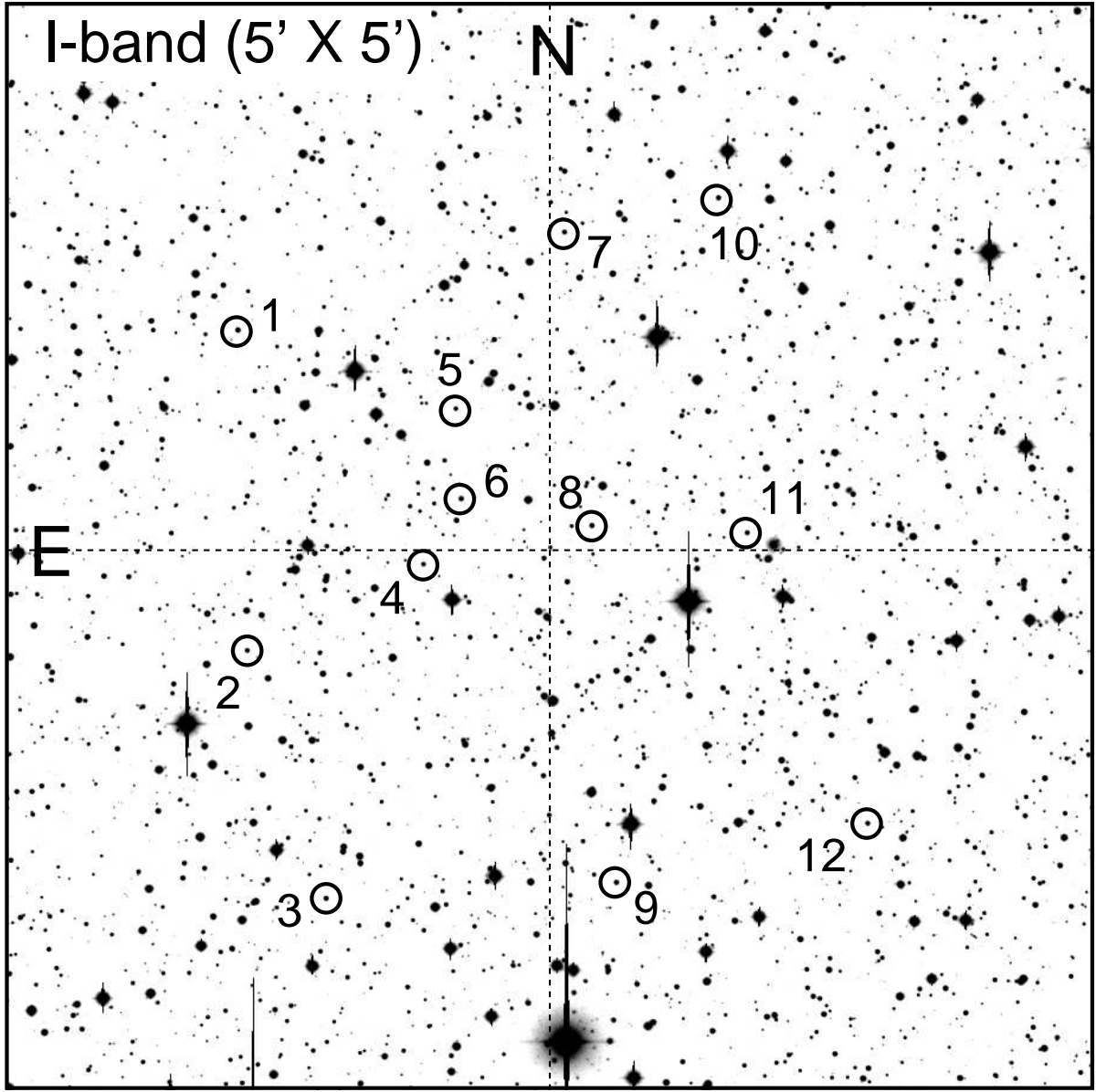}
\caption{
Wide-field (5\arcmin $\times$ 5\arcmin)
$I$-band image centered on \pmn.
The 12 reference
stars discussed in \S~\ref{sec:lco} are
circled and numbered. Magnitudes and
positions of these stars, relative to star \#8, are
provided in Table~\ref{tbl:photometry}.
}
\label{fig:field}
\end{figure}

\begin{figure}
\figurenum{5}
\plotone{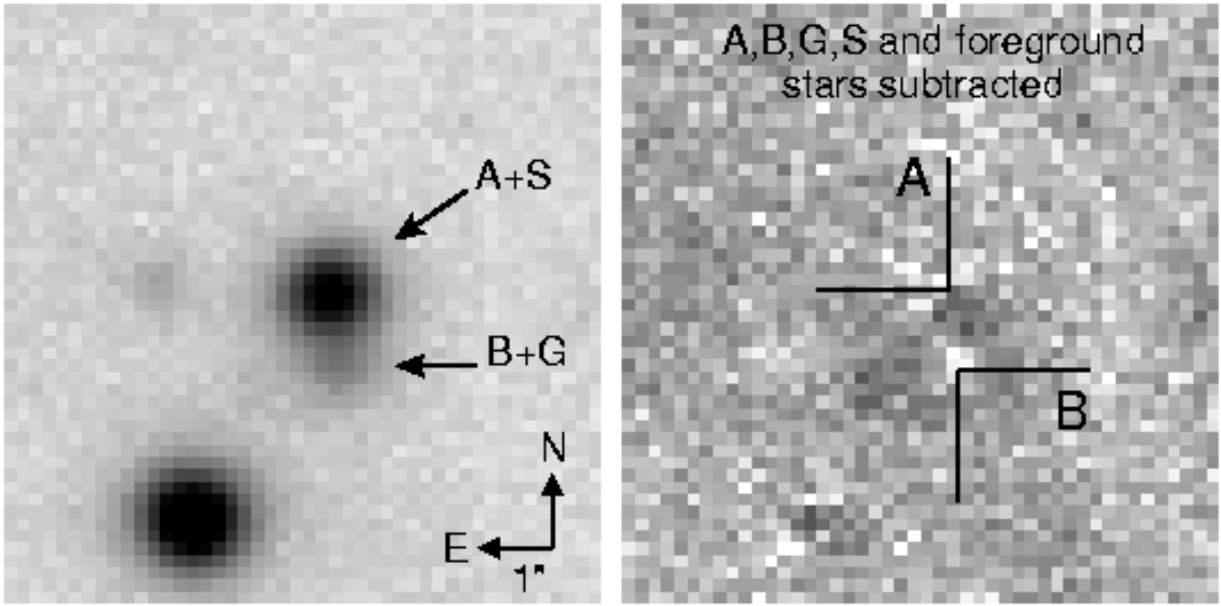}
\caption{
{\em Left.}
Ground-based $I$-band image
(8\arcsec $\times$ 8\arcsec) of \pmn.
Components A and S are unresolved,
as are components B and G.
There are foreground
stars to the southeast and east.
Gray levels are logarithmically scaled
from $-5\sigma$ (white)
to the central intensity of \pmn~(black).
{\em Right.}
Residual image after a
six-component model
describing \pmn~and the foreground stars
(see \S~\ref{sec:lco}) has been subtracted.
For clarity, only the positions of A and B are
marked.
Gray levels are linearly scaled
from $-5\sigma$ to $+5\sigma$. 
}
\label{fig:lco}
\end{figure}

During 1999 April 10-14 we obtained
$BRI$ images of \pmn~with the du~Pont~2.5m
telescope at Las Campanas Observatory (LCO).
We used the SITe\#3 $2048\times 4096$ CCD
camera, with gain 2.5 e$^{-}$/D.N.\ and read noise 6.6 e$^{-}$.
Table~\ref{tbl:lco} is a journal of these observations.
The images were bias-subtracted and flat-fielded with standard
IRAF\footnote{
IRAF is distributed by the National Optical Astronomy Observatories,
which are operated by the Association of Universities for Research
in Astronomy, Inc., under cooperative agreement with the National
Science Foundation.
}
procedures.
The rotation and pixel scale ($0\farcs1631$) of each image
were derived using
at least 30 stars from the USNO-A2.0 catalog \citep{usno}.
Figure~\ref{fig:field} shows the
$5\arcmin \times 5\arcmin$ $I$-band field.
The left panel of Figure~\ref{fig:lco} is an $8\arcsec \times 8\arcsec$
subraster centered on \pmn.

Photometry was performed with the DAOPHOT package
in IRAF. First, we constructed an empirical PSF
using the signal-weighted average of the images of
12 well-exposed, widely-spaced stars in the field.
These reference stars are circled and labeled in Figure~\ref{fig:field}.
Due to the crowded condition of the field, the PSF diameter
was limited to $5\arcsec$.
Next, we used this PSF template to fit simultaneously
for the positions and magnitudes of all 12 reference stars,
all the components representing \pmn, and all the neighbors
of the reference stars and \pmn~within $5\arcsec$.

For the $I$ and $R$ images the model of \pmn~consisted
of four point sources representing A, B, G, and S.
Their relative positions were fixed at the values
measured in the HST image.
For the $B$ images, which had poorer seeing
and a smaller signal, the covariance between
A and S and between B and G prevented convergence.
For these we used a model consisting of two components
with the same relative separation
as A and B in the HST image.

The right panel of Figure~\ref{fig:lco} shows
the $I$-band image at high contrast
after subtraction of the best-fit model.
The neighboring stars to the southeast and east
were also subtracted.
A pattern of positive residuals between A and B
at the $4\sigma$ level may be
unmodeled light from G. In all other
images the residuals were consistent with noise.

The instrumental magnitudes of the
reference stars and components of \pmn, relative
to reference star \#8,
are listed in Table~\ref{tbl:photometry}.
The quoted uncertainty in each magnitude difference
is the statistical error in the fit. For a
few stars in the $R$ and $B$ images the uncertainty
was increased to encompass the difference
between fits to the two different exposures.
We emphasize that the magnitudes of A and S are
covariant, as are those of B and G.
Components A and S were typically separated by just
one-third of the seeing disc and B and G by
only one-tenth.
For this reason, the combined magnitudes of A+S
and of B+G are also listed in Table~\ref{tbl:photometry}.

We connected the instrumental magnitudes to the
Johnson-Kron-Cousins photometric system
by observing at least one of the standard
fields described by \citet{landolt} each night.
Six stars in the field SA110-499 were used
to calibrate the $B$ and $R$ magnitude scales, using
the $B-R$ index to compute color terms.
Four stars in PG0918+029 and three stars in
PG1323--086 were used to calibrate the
$I$ magnitude scale, using the $R-I$ index
to compute color terms.
In all cases the aperture diameter was $14\arcsec$.
We adopted ``typical'' Las Campanas
extinction coefficients of
$k_I = 0.08$,
$k_R = 0.11$, and
$k_B = 0.28$.
The calibrated magnitudes of reference
star \#8 are printed
beneath Table~\ref{tbl:photometry}.
The quoted uncertainty is
the quadrature sum of the
uncertainty in the instrumental magnitude
and the RMS scatter in the calibration solution.
With this calibration,
the total apparent magnitudes of all
model components (A, B, S, and G)
during 1999 April 10-14 were
$I = 18.56$, $R = 19.16$, and $B = 20.48$,
within 0.04~mag.

The total $I$ magnitude disagrees with
the value derived from the HST/WFPC2 image ($I=18.79\pm 0.07$).
More significantly,
the LCO image exhibits a much larger contrast
between northern (A+S) and southern (B+G) components.
In the LCO image the flux ratio (A+S)/(B+G)
is $8.3 \pm 0.3$, whereas the WFPC2 value
is $4.6 \pm 0.4$. These discrepancies are possibly
the result of variability of the source quasar.
However, the significance of these
discrepancies is unclear, since the results were
derived from different instruments and photometric models.
Repeated ground-based measurements will be important
in assessing variability, using Table~\ref{tbl:photometry}
as a baseline.

\section{Optical spectroscopy}
\label{sec:spectra}

Optical spectra of the quasar and lens galaxy
were obtained with the 3.6m telescope of
the European Southern Observatory (ESO) at La Silla, Chile.
We used the ESO Faint Object Spectrograph
and Camera (EFOSC) with CCD \#40 ($2048 \times 2048$,
gain 1.3 e$^{-}$/D.N., readout noise 7.5 e$^{-}$) and
grism \#6 (3860-8070\AA, 30~gr/mm) giving a plate scale
of 2\AA~per pixel.
We obtained two separate spectra, one in which
the slit was centered on quasar component A
and one centered on the lens galaxy G.
In both cases the slit was $0\farcs7$ wide and
oriented east-west in order to isolate the
components as much as possible.

\begin{figure}
\figurenum{6}
\plotone{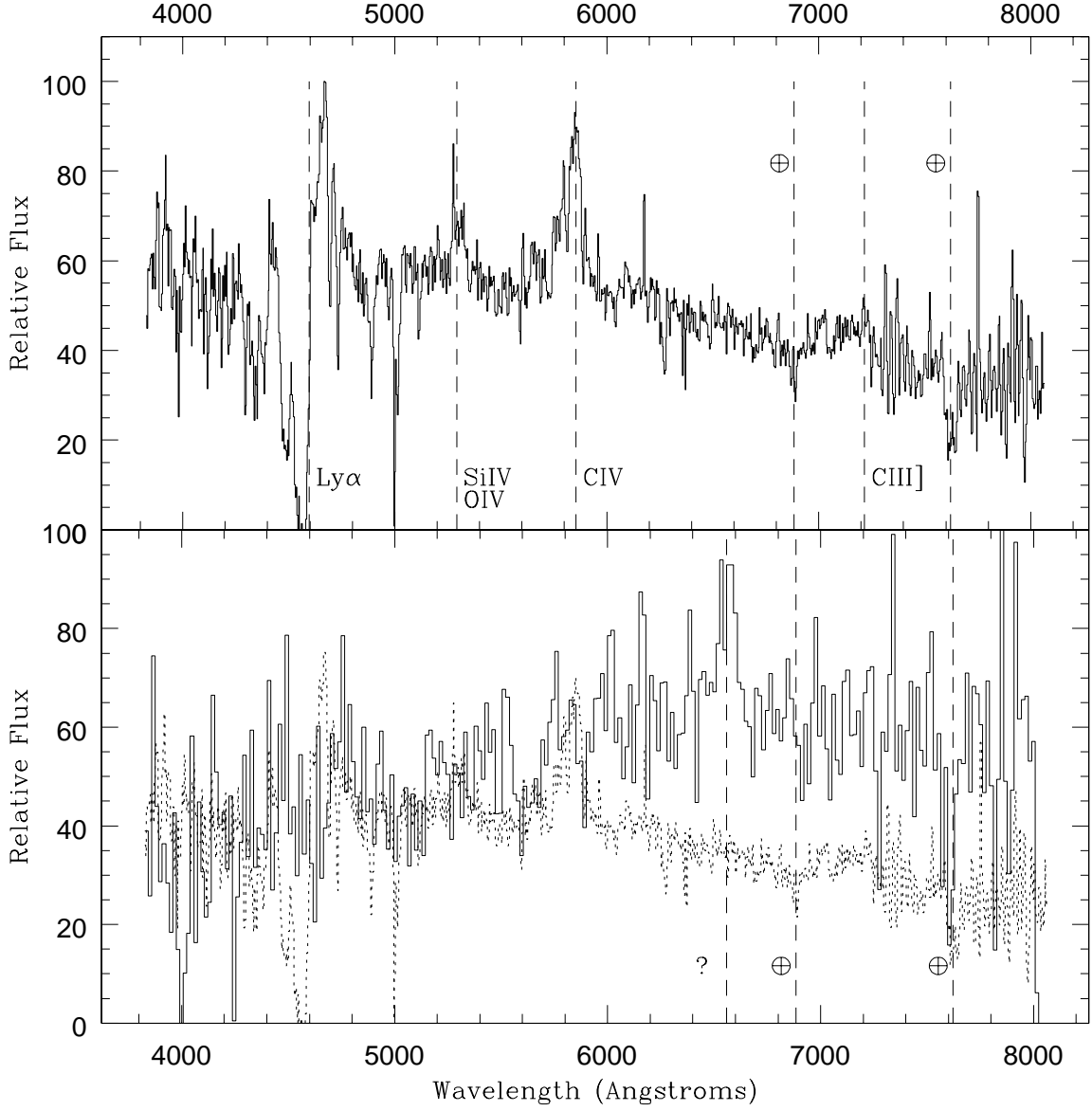}
\caption{
{\em Top.}
Optical spectrum of quasar component A
with resolution 10\AA.
Dashed lines indicate
emission lines at $z=2.78$. Telluric lines
($\oplus$) are also marked.
{\em Bottom.}
Optical spectrum of the lens galaxy
with resolution 30\AA.
The dotted line is the quasar spectrum
from the top panel, after scaling by the B/A
flux ratio and adjusting for exposure time.
Due to variable seeing and non-photometric
conditions the scale factor is uncertain
by at least 15\%.
}
\label{fig:spectra}
\end{figure}

The quasar spectrum is shown in the top panel of
Figure~\ref{fig:spectra}.
It was obtained
on 2000 March 4 in a 30-minute exposure in $1\farcs8$ seeing,
through an airmass of 1.5 and has a resolution of 10\AA.
No attempt was made to correct for differential atmospheric refraction.
Three broad emission lines are obvious,
corresponding to Ly$\alpha$ (1216\AA), \ion{Si}{4}+\ion{O}{4} (1400\AA),
and \ion{C}{4} (1549\AA) at
a redshift of $2.78 \pm 0.01$. This redshift was determined
with reference to the \ion{Si}{4}+\ion{O}{4} and \ion{C}{4} lines only,
since the broad absorption trough blueward of Ly-$\alpha$
makes centroid estimation difficult.
Once this redshift was established, a probable emission
line of \ion{C}{3}] (1909\AA) was identified.

At least two significant
non-terrestrial absorption lines are also present
in the spectrum (4890\AA\ and 5000\AA),
but we are unable to assign them confidently
to systems at particular redshifts.
We did not identify any $z=0$ stellar 
absorption lines in this spectrum, which would have
confirmed the hypothesis of \S~\ref{sec:hst} that component S
is a foreground Galactic star.
This is not surprising, because
the A/S flux ratio is 8.1 in the HST image
while the signal-to-noise ratio
of each resolution element is only 7.5.

The lens galaxy spectrum, shown in the bottom panel of
Figure~\ref{fig:spectra}, was obtained on 2000 March 6 with
a 60-minute exposure through an average airmass of 1.5 in
1\farcs3 seeing.
The resolution has been degraded to 30\AA\ by
combining wavelength bins, in order to
boost the signal-to-noise.
The slit also admitted
the southern quasar component B, but based on the
HST photometric model (Table~\ref{tbl:hst})
the light from G was expected to dominate B.

There is only one definite
feature, an emission line at 6560\AA\ with
a S/N of about 3.
This feature does not appear in the quasar spectrum,
which has been scaled to the level expected from
the B component and plotted as a dotted line in
Figure~\ref{fig:spectra}.
Although the wavelength of this feature is
suspiciously close to $z=0$ H$\alpha$,
suggesting it might result from
imperfect background subtraction, there were
several brighter sky lines that
were subtracted successfully.
It is therefore tempting to identify this
feature with emission from the lens galaxy.

With one line of unknown origin,
redshift determination is impossible.
However, it is possible to estimate the lens
galaxy redshift from photometry alone,
by requiring the photometric properties to be consistent
with the passively-evolving fundamental plane (FP)
of early-type galaxies. \citet{fundplane}
described the FP method
and applied it to 17 lenses with known spectroscopic
redshifts, finding a scatter of 0.11 between the FP
and spectroscopic redshifts.
Applied to \pmn, the lens galaxy redshift estimate
is $z_{FP} = 0.36 \pm 0.08$.
Given this redshift estimate,
one plausible identification
for the emission at 6560\AA\ is [\ion{O}{3}] (5007\AA)
at $z=0.31$. The emission line appears to
be resolved, which could be a result of blending
with [\ion{O}{3}] (4959\AA) and/or H$\beta$ (4861\AA).
In this case one would
expect to see these lines in a deeper spectrum, along
with [\ion{O}{2}] (3727\AA) at 4882\AA.
A redder spectrum would reveal H$\alpha$ (6563\AA) at
8598\AA.

Obviously there are many other possibilities,
if the lens redshift is not correctly predicted
by the FP method. Nevertheless, for the purpose of lens modeling
(\S~\ref{sec:model}) we adopt $z_l = 0.36$ as a
working hypothesis.

\section{Models of the lens potential}
\label{sec:model}

The present observations of \pmn~provide
only five useful
constraints on the lens potential:
the positions of B and G relative to A
in the HST/WFPC2
image (Table~\ref{tbl:hst})
and the A/B radio flux density ratios
(Table~\ref{tbl:vla}).
The optical magnitude
differences are not useful because they
are affected by dust extinction, reddening
and possibly microlensing in the lens galaxy.
Any additional lensed images
must be dimmer than A by a factor of at least
200, based on the radio images.
We therefore confine ourselves to simple two-image models.

Our first purpose is to confirm
that lensing is a natural explanation for the
observed image configuration.
For this purpose we consider the
simplest plausible model for the
dark-matter halo of a galaxy,
an isothermal sphere. This model produces
two images with a magnification ratio equal
to the ratio of their distances from the
lens center. The lens galaxy center
in the HST/WFPC2
image is displaced 8 mas east of the
line joining A and B (see Table~\ref{tbl:hst});
this is within the positional uncertainty,
so an isothermal sphere is a viable model.

The image separations in Table~\ref{tbl:hst}
imply a magnification ratio of $11.8\pm 1.7$.
This compares well with the
radio flux density ratios in Table~\ref{tbl:vla},
although it is only in actual agreement with
the VLBA ratio.
There is a good reason why one might expect the VLBA
flux density ratio to correspond more
closely to the magnification ratio: it may exclude
relatively diffuse flux that is lumped
into component A in the VLA images (see \S~\ref{sec:vlba}).
Another possible explanation for the small but significant
discrepancies is source variability. The proper comparison
to the magnification ratio is not the instantaneous
flux density ratio,
but rather the ratio of light curves
that have been shifted
by the appropriate time delay.

Our second purpose is to
model the system in the same manner as
other two-image lenses observed
in the CASTLES program \citep{lehar},
for the sake of uniformity and
to provide the starting point for
more detailed models.
Although the rough agreement with the
isothermal sphere model is encouraging,
it would be naive to conclude
that it is an adequate model for the lens potential.
Numerous studies have shown that,
while an isothermal sphere
provides a good first approximation,
extra terms representing
internal or external shear must be added
at the 10-20\% level to
match all the observational constraints \citep{kks}.
We therefore consider a singular isothermal
ellipsoid (SIE), in which the
surface mass density is (in units
of the critical surface density for lensing),
\begin{equation}
\kappa(x,y) = \frac{b}{2}\left[ \frac{2q^2}{1+q^2} \left(x^2 + y^2/q^2\right) \right]^{-1/2},
\end{equation}
where $q$ is the axis ratio and $b$ sets the mass
scale.

There are five parameters ($b$ and $q$ as above, and three
implied parameters for the position angle of the axes and
the source coordinates). Since there are
also five constraints, the parameters are determined uniquely.
We chose the mean 
of the radio flux density
ratios listed in Table~\ref{tbl:vla}
as the constraint on the magnification
ratio, with the caveats discussed above.
We determined parameter uncertainties by surveying
the range of parameters for which $\Delta\chi^2 < 1$.
The results are listed in Table~\ref{tbl:lensmodel}.

We used this model to predict the quantity
$h\Delta t$, where $\Delta t$ is the time delay
and $H_0 = 100h$~km/s/Mpc.
The prediction depends on the lens galaxy redshift,
the cosmological parameters $\Omega_m$ and
$\Omega_{\Lambda}$, and the clumpiness of
matter on cosmological scales.
For the lens galaxy redshift we adopt
the FP estimate $z_l = 0.36$ (see \S~\ref{sec:spectra}).
Under the further assumptions
$\Omega_m = 0.3$ and $\Omega_\Lambda = 0.7$,
and using the filled-beam approximation,
$h\Delta t = 14.9 \pm 0.2$ days.
If instead
$\Omega_m = 1$ and $\Omega_\Lambda = 0$
then $h\Delta t = 14.0 \pm 0.2$ days.
The quoted uncertainties are based on
the SIE model only, and are
therefore underestimates; many authors
have shown that a wider class
of lens models needs to be explored
to obtain realistic error
estimates (e.g.\
\cite{koch91,bernstein,williams}).

\section{Summary and future prospects}
\label{sec:summary}

We now summarize the argument that \pmn~is
a gravitational lens.
It consists of two flat-spectrum radio components,
each of which is compact
on milliarcsecond scales and has a stellar
optical counterpart. This implies both
components are quasars, and indeed the bright component
is a spectroscopically verified $z=2.78$ quasar.
Radio-loud quasars are scarce enough that
when two of them are observed within
$1\arcsec$, they are probably
either a physical binary or a pair of lensed images.
That the spectral indices of each component are the same
suggests lensing is the explanation. A high-resolution
optical image shows
diffuse emission (the lens galaxy) along the line
between the components. The image configuration
is consistent with
the simplest plausible mass distribution
for the lens galaxy.

Will this new lens be useful in the enterprise of using
time delays to constrain cosmological parameters and/or
aspects of galaxy structure? The answer depends on
whether the time delay of \pmn~can be measured
and whether more constraints on the potential of the lens
galaxy are likely to be discovered.

As for the first issue, measuring the time delay,
prospects are good.
There are several indications that \pmn~is variable at
radio wavelengths. Its spectral index
is flat and the components are compact
on milliarcsecond scales, both of which are
indicators of variability. In addition, the three
measurements of total flux density at 8.5~GHz
listed in Table~\ref{tbl:fluxes} (two from the VLA
and one from ATCA) are all significantly different.
At optical wavelengths we do not
have as much information to judge variability.
The discrepancy between the ground-based (LCO)
$I$-band photometry and the WFPC2
photometry (see \S~\ref{sec:lco}) is one
suggestion of variability, but further
monitoring is required.
The local reference system established
in \S~\ref{sec:lco} and Table~\ref{tbl:photometry}
will be of use in future measurements.

The second issue, constraining
the lens potential, is the greater challenge.
At least the lens appears to be a single galaxy, and
the lens position and magnification ratio are
consistent with the simplest plausible models, giving
\pmn~an advantage over more obviously
complicated lenses with measured time delays,
such as Q0957+561 and CLASS~B1608+656.
The challenge will be to obtain enough constraints
to explore a wider class of models.
Deeper or higher-resolution VLBI observations
may reveal corresponding substructure
in the lensed images.
Infrared images from space or with
adaptive optics will improve estimates for
the lens galaxy's position and shape. They would
also clarify the nature of the
host galaxy that was tentatively identified in \S~\ref{sec:hst},
which could provide very important model constraints \citep{rings}.
Obtaining the lens galaxy redshift is also a priority.
If these observational challenges
can be met, \pmn~will contribute to the
growing field of using lensing phenomena
to study galaxy structure and cosmology.

\acknowledgments

We are grateful to Alok Patnaik
and Andr\'{e} Fletcher
for their help with the lens search.
We thank Joan Wrobel,
Geoff Bower and
Greg Taylor
for assistance with NRAO telescopes.
Chris Fassnacht and Steve Myers kindly
made their VLA mapping script available.
John Tonry provided valuable advice regarding spectroscopy.
This research was supported by the National Science Foundation
under grants AST-9617028 and AST-9616866.
M.T.R.\ received partial support from Fondecyt grant 1980659
and Catedra Presidencial en Ciencias.
J.N.W.\ thanks the
Fannie and John Hertz foundation
for financial support.

\clearpage

\clearpage

\clearpage

\clearpage

\clearpage

\clearpage

\clearpage

\begin{deluxetable}{lcccccccc}
\rotate

\tabletypesize{\scriptsize}
\tablecaption{Photometric models based on VLA and VLBA data\label{tbl:vla}}
\tablewidth{0pt}

\tablehead{
\colhead{Date}              &
\colhead{Frequency}         &
\colhead{Beam FWHM} &
\multicolumn{2}{c}{Flux density}      &
\colhead{RMS noise}         &
\colhead{$\Delta$R.A.}      &
\colhead{$\Delta$Decl.}     &
\colhead{Flux density}        \\

\colhead{} &
\colhead{(GHz)} &
\colhead{(mas $\times$ mas, P.A.)} &
\colhead{A (mJy)} &
\colhead{B (mJy)} &
\colhead{(mJy/beam)} &
\colhead{(mas)} &
\colhead{(mas)} &
\colhead{ratio}
}

\startdata
19 May 1998 & 8.46 & $520 \times 180$ (0\degr)    & 192.0  & 13.7  & 0.35 & $-94 \pm 3$  & $-985  \pm 7$  & $13.97 \pm 0.35$ \\
19 Jul 1999 & 8.46 & $590 \times 210$ ($-1$\degr) & 169.7  & 11.5  & 0.24 & $-94 \pm 3$  & $-998  \pm 6$  & $14.73 \pm 0.31$ \\
19 Jul 1999 & 4.86 & $960 \times 340$ (2\degr)    & 200.4  & 13.2  & 0.20 & $-88 \pm 3$  & $-1013 \pm 8$  & $15.22 \pm 0.23$ \\
19 Jul 1999 & 14.94& $300 \times 110$ (7\degr)    & 168.8  & 11.7  & 0.49 & $-97 \pm 3$  & $-987  \pm 7$  & $14.45 \pm 0.61$ \\
11 Oct 1999 &4.975 & $10.5\times 1.8$ ($-16$\degr)& 145.4  & 13.7& 0.19& $-96.99\pm 0.04$&$-991.32\pm 0.14$& $10.61 \pm 0.14$ 
\enddata

\tablecomments{
The first four rows are based on VLA data (\S~\ref{sec:vla}).
The last row is based on VLBA data (\S~\ref{sec:vlba}).
The RMS level in each image was taken as
an estimate of the uncertainty in the relative flux density scale.
The absolute flux density scale is uncertain by an additional 3\%
for the first 3 rows and 5\% for the last 2 rows, based
on VLA documentation.
The uncertainty in each coordinate
was estimated as the beam FWHM
divided by twice the S/N (peak/RMS) of the component. 
}

\end{deluxetable}

\begin{deluxetable}{ccccc}
\tabletypesize{\scriptsize}
\tablecaption{Two-point spectral indices based on VLA data\label{tbl:indices}}
\tablewidth{0pt}

\tablehead{
\colhead{Frequency 1} &
\colhead{Frequency 2} &
\colhead{Component A} &
\colhead{Component B} &
\colhead{Uncertainty due to} \\

\colhead{(GHz)} &
\colhead{(GHz)} &
\colhead{} &
\colhead{} &
\colhead{absolute flux scale}
}

\startdata
4.86 & 8.46  & $-0.299 \pm 0.003$ & $-0.24 \pm 0.05$ & 0.08 \\
8.46 & 14.94 & $-0.009 \pm 0.006$ & $+0.02 \pm 0.08$ & 0.10 \\ 
4.86 & 14.94 & $-0.152 \pm 0.003$ & $-0.11 \pm 0.04$ & 0.05
\enddata
\tablecomments{
Spectral indices $\alpha$ are defined such that
$S_{\nu} \propto \nu^{\alpha}$. 
Uncertainties quoted in columns 3 and 4
derive only from the RMS level in each image.
Column 5 reports the additional uncertainty
due to the absolute flux scales, which
affects A and B identically.
}

\end{deluxetable}

\begin{deluxetable}{lcclll}
\tabletypesize{\scriptsize}
\tablecaption{Compilation of radio flux density measurements\label{tbl:fluxes}}
\tablewidth{0pt}

\tablehead{
\colhead{} &
\colhead{} &
\colhead{Frequency}  &
\multicolumn{3}{c}{Flux density} \\

\colhead{Date} &
\colhead{Observatory} &
\colhead{(GHz)} &
\colhead{A (mJy)} &
\colhead{B (mJy)} &
\colhead{A+B (mJy)}
}

\startdata
1974-83     & UTRAO\tablenotemark{a}         & 0.365 & \nodata & \nodata & $451   \pm 36  $ \\
May 1996    & VLA (DnC config.)\tablenotemark{b} & 1.40  & \nodata & \nodata & $280.2 \pm  8.4$ \\
1979        & Parkes\tablenotemark{c}        & 2.70  & \nodata & \nodata & $240           $ \\
Nov 1990    & Parkes\tablenotemark{d}        & 4.85  & \nodata & \nodata & $258   \pm 21  $ \\
19 Jul 1999 & VLA (A config.)                  & 4.86  & 200.4   & 13.2    & $213.6 \pm 6.4 $ \\
11 Oct 1999 & VLBA & 4.975 & 145.4\tablenotemark{e} & 13.7    & $175.9 \pm 8.8\tablenotemark{f} $ \\
5 Apr 2000  & ATCA (6D config.)               & 4.80  & \nodata & \nodata & $218.7 \pm 6.8 $ \\
19 May 1998 & VLA (A config.)                  & 8.46  & 192.0   & 13.7    & $205.7 \pm 6.2 $ \\
19 Jul 1999 & VLA (A config.)                  & 8.46  & 169.7   & 11.5    & $181.2 \pm 5.4 $ \\
5 Apr 2000  & ATCA (6D config.)               & 8.64  & \nodata & \nodata & $264.2 \pm 8.6 $ \\
19 Jul 1999 & VLA (A config.)                  & 14.94 & 168.8   & 11.7    & $180.5 \pm 9.0 $ 
\enddata

\tablenotetext{a}{
University of Texas Radio Astronomy Observatory \citep{texas}.
The Texas position differs from our position by
103\arcsec~in right ascension. We believe this is due to lobeshift,
because the entry is flagged as
possibly lobeshifted and the lobeshift increment is 52\arcsec.
}

\tablenotetext{b}{
NRAO VLA Sky Survey \citep{nvss}.
}

\tablenotetext{c}{
Parkes Catalog, PKSCAT90 \citep{pkscat}.
}

\tablenotetext{d}{
Parkes-MIT-NRAO zenith catalog \citep{pmnz}.
}

\tablenotetext{e}{
Flux density of A does not include the diffuse emission to the west.
}

\tablenotetext{f}{
Total flux density does include the diffuse emission west of A.
}

\tablecomments{
For observations that could not resolve A and B,
only the total flux density is reported.
}

\end{deluxetable}

\begin{deluxetable}{ccccc}
\tabletypesize{\scriptsize}
\tablecaption{Photometric model based on HST/WFPC2 image\label{tbl:hst}}
\tablewidth{0pt}

\tablehead{
\colhead{Component} &
\colhead{$\Delta$R.A.} &
\colhead{$\Delta$Decl.} &
\colhead{Relative} &
\colhead{$R_{{\mathrm e}{\mathrm f}{\mathrm f}}$} \\

\colhead{} &
\colhead{(mas)} &
\colhead{(mas)} &
\colhead{flux} &
\colhead{(arcsec)} 
}

\startdata
A &        0     &        0     & $100 \pm 6   $ & \nodata \\
B & $-101 \pm 6$ & $-987 \pm 5$ & $3.77\pm 0.55$ & \nodata \\
G & $-85 \pm 6$ & $-911 \pm 6$& $20.9 \pm 1.3$ & 0.20    \\
S & $ 239 \pm 2$ & $-140 \pm 5$ & $12.3 \pm 1.6$ & \nodata \\
p & $ 70 \pm 11$ & $ 81 \pm 17$ & $5.7 \pm 2.5$ & \nodata \\
q & $104 \pm 8$  & $-37 \pm 21$ & $4.7 \pm 0.8$ & \nodata
\enddata

\end{deluxetable}

\begin{deluxetable}{lcccc}
\tabletypesize{\scriptsize}
\tablecaption{Journal of ground-based optical observations\label{tbl:lco}}
\tablewidth{0pt}

\tablehead{
\colhead{Date} &
\colhead{Filter} &
\colhead{Duration} &
\colhead{Seeing} &
\colhead{Airmass} \\

\colhead{} &
\colhead{} &
\colhead{(sec)}&
\colhead{(arcsec)} &
\colhead{}
}

\startdata
10 April 1999 & $R$ & 600  & 0.81 &  1.02 \\
11 April 1999 & $R$ & 1000 & 0.88 &  1.01 \\
11 April 1999 & $B$ & 1000 & 0.92 &  1.01 \\
12 April 1999 & $I$ & 1500 & 0.63 &  1.01 \\
14 April 1999 & $B$ & 1800 & 0.88 &  1.01
\enddata

\end{deluxetable}

\begin{deluxetable}{lrrrrr}
\tabletypesize{\scriptsize}
\tablecaption{Photometry and astrometry of \pmn~and 12 reference stars\label{tbl:photometry}}
\tablewidth{0pt}

\tablehead{

\colhead{Object} &
\colhead{$\Delta$R.A.} &
\colhead{$\Delta$Decl.} &
\colhead{$\Delta I$} &
\colhead{$\Delta R$} &
\colhead{$\Delta B$} \\

\colhead{}   &
\colhead{(sec)}    &
\colhead{(arcsec)} &
\colhead{(mag)}  &
\colhead{(mag)}  &
\colhead{(mag)}

}

\startdata
A    & $  0.925$ & $ -5.96$ & $ 1.075(8)  $ & $ 1.002(31) $ & \nodata       \\
S    & \nodata   & \nodata  & $ 3.238(51) $ & $ 4.252(129)$ & \nodata       \\
G    & \nodata   & \nodata  & $ 3.616(220)$ & $ 4.009(264)$ & \nodata       \\
B    & \nodata   & \nodata  & $ 4.559(514)$ & $ 4.889(116)$ & \nodata       \\
A+S  & \nodata   & \nodata  & $ 0.936(8)  $ & $ 0.948(31) $ & $ 0.991(4)  $ \\
B+G  & \nodata   & \nodata  & $ 3.236(220)$ & $ 3.586(264)$ & $ 4.383(53) $ \\
1    & $  7.866$ & $ 53.72$ & $-0.126(4)  $ & $ 0.667(4)  $ & $ 2.192(9)  $ \\
2    & $  7.663$ & $-34.34$ & $ 0.261(4)  $ & $ 0.259(3)  $ & $ 0.433(4)  $ \\
3    & $  5.913$ & $-102.50$& $ 0.071(3)  $ & $ 0.057(4)  $ & $ 0.132(3)  $ \\
4    & $  3.745$ & $-10.64$ & $ 0.447(3)  $ & $ 0.512(5)  $ & $ 0.688(4)  $ \\
5    & $  3.026$ & $ 32.13$ & $ 0.662(4)  $ & $ 0.748(4)  $ & $ 1.076(4)  $ \\
6    & $  2.906$ & $  7.41$ & $-0.021(3)  $ & $ 0.093(4)  $ & $ 0.451(4)  $ \\
7    & $  0.608$ & $ 80.69$ & $ 0.259(4)  $ & $ 0.291(4)  $ & $ 0.528(4)  $ \\
8    & $  0.000$ & $  0.00$ & $ 0.000(0)  $ & $ 0.000(0)  $ & $ 0.000(0)  $ \\
9    & $ -0.543$ & $-98.09$ & $ 0.043(4)  $ & $ 0.139(4)  $ & $ 0.564(4)  $ \\
10   & $ -2.813$ & $ 90.13$ & $ 0.069(4)  $ & $ 0.214(4)  $ & $ 0.595(4)  $ \\
11   & $ -3.451$ & $ -1.95$ & $ 0.277(4)  $ & $ 0.246(3)  $ & $ 0.179(4)  $ \\
12   & $ -6.127$ & $-81.81$ & $ 0.219(4)  $ & $ 0.283(3)  $ & $ 0.556(4)  $
\enddata

\tablecomments{
Position and magnitude differences
are computed in the sense $x_{n} - x_{8}$.
Uncertainties in millimagnitudes are
contained in parentheses.
The coordinates of star \#8 are
R.A.~(J2000)$~=
18^{{\mathrm h}}38^{{\mathrm m}}27^{{\mathrm s}}.59$,
Dec.~(J2000)~$ =
-34\arcdeg 27\arcmin 35\farcs7$
within 0\farcs2.
The calibrated magnitudes of star \#8 are
$I = 17.75 \pm 0.02$, $R = 18.30 \pm 0.04$, $B = 19.54 \pm 0.04$.
} 

\end{deluxetable}

\begin{deluxetable}{ccccccc}
\tabletypesize{\scriptsize}
\tablecaption{Singular isothermal ellipsoid lens model\label{tbl:lensmodel}}
\tablewidth{0pt}
\tablehead{
  $x$ (\arcsec~East)
 &$y$ (\arcsec~North)
 &$b$ (\arcsec)
 &$\epsilon$ (1--$b/a$)
 &$\theta_\epsilon$
}
\startdata
%  x (''east)        y (''north)          b ('')            elip       theta_elip 
 $-0.085\pm0.006$ & $-0.911\pm0.006$ & $0.51\pm0.01  $ & $0.15\pm0.13$ & $19\pm32$ \\
\enddata

\tablecomments{
All coordinates are relative to the position of quasar component A.
In this model, the unlensed source is at ($-0\farcs033$, $-0\farcs498$)
and its flux is 40.2\% that of A.
}
\end{deluxetable}

\end{document}